\documentclass[sigconf,screen,nonacm,9pt]{acmart}

\usepackage{lmodern}
\usepackage{amsmath}

\usepackage{amssymb}
\usepackage{graphicx}
\usepackage{booktabs}
\usepackage{multirow}
\usepackage{xcolor}
\usepackage{tikz}
\usepackage{pgfplots}
\usepackage{pgfplotstable}
\usepackage{enumitem}
\usepackage{placeins}
\pgfplotsset{compat=1.18}
\usetikzlibrary{shapes.geometric,arrows.meta,positioning,calc,backgrounds,patterns}
\definecolor{pastelblue}{RGB}{174,198,232}
\definecolor{pastelorange}{RGB}{255,190,122}
\definecolor{pastelgreen}{RGB}{152,223,138}
\definecolor{pastelred}{RGB}{255,152,150}
\definecolor{pastelpurple}{RGB}{197,176,213}
\definecolor{pastelteal}{RGB}{158,218,229}
\definecolor{deepblue}{RGB}{31,119,180}
\definecolor{deeporange}{RGB}{255,127,14}
\definecolor{deepgreen}{RGB}{44,160,44}
\definecolor{deepred}{RGB}{214,39,40}
\definecolor{deeppurple}{RGB}{148,103,189}



\title{Phase Matters: Characterizing Heterogeneous Vision-Language Inference on a Mobile SoC}

\author{Aryama V Murthy}
\authornote{Equal contribution.}
\affiliation{\institution{International Institute of Information Technology, Hyderabad}\city{Hyderabad}\country{India}}
\author{Yashas N Kotre}
\authornotemark[1]
\affiliation{\institution{International Institute of Information Technology, Hyderabad}\city{Hyderabad}\country{India}}
\author{Prathmesh Sharma}
\authornotemark[1]
\affiliation{\institution{International Institute of Information Technology, Hyderabad}\city{Hyderabad}\country{India}}
\author{Pragya Mishra}
\authornotemark[1]
\affiliation{\institution{International Institute of Information Technology, Hyderabad}\city{Hyderabad}\country{India}}
\author{Sanjith Ganapathi}
\affiliation{\institution{International Institute of Information Technology, Hyderabad}\city{Hyderabad}\country{India}}
\author{Priyesh Shukla}
\affiliation{\institution{International Institute of Information Technology, Hyderabad}\city{Hyderabad}\country{India}}

\begin{document}

\acmConference[arXiv preprint]{arXiv preprint}{June 22, 2026}{}

\begin{abstract}
Recent phone-class mobile SoCs expose practical NPU execution paths for
on-device vision-language model (VLM) inference, but developers still lack
phase-level guidance for mapping VLM pipelines across heterogeneous backends.
We present a hardware-in-the-loop characterization of VLM inference on the
Qualcomm SM8750 (Snapdragon 8~Elite), covering phase throughput, cache-state
effects, 100-run thermal stability, energy, heterogeneous CPU/NPU pipeline
configurations, and visual-token-budget sensitivity. Using FastVLM-0.5B as an
end-to-end case study, together with encoder-only measurements across four
architecture families, we show that \emph{phase matters}: NPU execution is
highly phase-dependent, delivering \textbf{1.64$\times$} speedup for prefill
but only \textbf{1.18$\times$} for decode, while vision encoders achieve
\textbf{20--45$\times$} speedups over CPU. These gains translate into
\textbf{10.47$^\circ$C} lower steady-state temperature and
\textbf{2.52$\times$} lower energy, avoiding thermal throttling in always-on
settings. Finally, we show that a four-step graph rewrite enables previously
unsupported encoders, such as Phi-3.5-V, to reach the QNN path with up to
\textbf{22$\times$} speedup, providing a practical porting recipe for mobile
VLM deployment.
\end{abstract}

\maketitle
\pagestyle{empty}

\section{Introduction}
\label{sec:intro}

Vision-language models (VLMs) are crossing a critical threshold: devices already
in consumers' hands---with billion-parameter multimodal models---can now run
VLM inference entirely on-device, with no cloud round-trip.
Wearables, AR overlays, robotics, and privacy-sensitive pipelines all demand
exactly this capability~\cite{fastvlm,mobilevlm,litevlm}.
The Qualcomm SM8750 (Snapdragon 8~Elite) exposes a stable Hexagon NPU path
for full multimodal models via the QNN SDK~\cite{qualcomm_qnn,qualcomm_genai},
making it the right deployment platform to ask:
\emph{what does NPU acceleration concretely deliver for each phase of VLM
inference, and how should practitioners configure heterogeneous pipelines?}

This question is non-trivial: VLM inference decomposes into three stages with
fundamentally different hardware affinities~\cite{llmnpu}.
\emph{Prefill} saturates NPU tensor engines with batched GEMMs.
\emph{Decode} is bandwidth-bound and shape-dynamic, favouring CPU
cache locality~\cite{heteroinfer,qualcomm_genai}.
\emph{Vision encoding} is static-shape and compute-dense---the NPU's clearest
sweet spot, yet the least systematically studied.
Treating these phases uniformly leads to measurable waste in latency, energy,
and thermal budget.
Prior systems work on heterogeneous mobile inference targets text-only
LLMs~\cite{llmnpu,heteroinfer}; the vision encoding phase, its thermal
behavior, and practical porting paths for unsupported encoders are not
covered.
We address this gap with a VLM-specific, hardware-in-the-loop study on SM8750,
contributing phase-aware deployment guidance and a practical QNN porting
workflow for production phone NPUs.

\textbf{Contributions:}
\begin{itemize}[leftmargin=*,topsep=1pt,itemsep=1pt,parsep=0pt]
\item \textbf{Phase map:} 1.64$\times$ NPU prefill, 1.18$\times$ decode,
  6.41$\times$ cold-start overhead (Sec.~\ref{sec:e2e}).
\item \textbf{Thermal suite:} 10.47$^\circ$C cooler, 2.52$\times$ lower
  energy, zero throttling on NPU over 100 runs (Sec.~\ref{sec:thermal}).
\item \textbf{Pipeline map:} Five CPU$\leftrightarrow$NPU configurations;
  W4A8 hybrid at 97.5\,tok/s (Sec.~\ref{sec:pipeline}).
\item \textbf{Latency knob:} Token-budget sweep 0.36--0.85\,s TTFS, no
  recompilation (Sec.~\ref{sec:pipeline}).
\item \textbf{Encoder scaling:} Four families $\times$ CPU/GPU/NPU,
  20--45$\times$ CPU-to-NPU speedups (Sec.~\ref{sec:encoders}).
\item \textbf{Porting methodology:} Four-step rewrite, 22$\times$ on
  Phi-3.5-V (Sec.~\ref{sec:rewrite}).
\end{itemize}

\section{Platform and Experimental Setup}
\label{sec:setup}

\subsection{Hardware: Qualcomm SM8750 Snapdragon 8 Elite}

Our target is the Qualcomm SM8750~\cite{snapdragon_brief}, a 3\,nm SoC with
a 64-bit Oryon CPU (4.32\,GHz), LPDDR5x at 5300\,MHz, Adreno GPU, and a
Hexagon NPU with scalar/vec\-tor/ten\-sor engines and VTCM
scratch~\cite{qualcomm_genai}.
W4A8 INT4-weight packing halves VTCM bandwidth demand vs.\ INT8~\cite{w4a8,awq}.
Static-shape NPU paths use QNN SDK~\cite{qualcomm_qnn} via ONNX; dynamic
delegation uses LiteRT~\cite{litert}.

\subsection{VLM Phase Decomposition}

Following the roofline analysis framework~\cite{williams2009roofline}, VLM
inference separates into three hardware-distinct phases (Table~\ref{tab:phases}).
Prefill has high operational intensity $I = \text{FLOPs}/\text{Byte}$ due to
batched GEMM and ViT attention over full image and prompt---NPU systolic engines
achieve peak utilization here~\cite{qualcomm_genai,llmnpu}.
Decode processes one token per step while reading a growing KV cache:
per-layer traffic scales as $2LDb$ bytes, making it bandwidth-bound.
The NPU delivers 1.18$\times$ decode throughput; heterogeneous pipelines
assign decode to CPU to maximize energy efficiency and preserve NPU capacity
for the compute-intensive prefill phase~\cite{heteroinfer}.
Vision encoding is entirely static-shape and compute-dense, yielding the
highest NPU efficiency of any VLM stage~\cite{fastvlm,clip,siglip}.

\begin{table}[t]
\centering
\caption{VLM inference phase decomposition and backend affinity}
\label{tab:phases}
\scriptsize
\setlength{\tabcolsep}{4pt}
\begin{tabular}{@{}llll@{}}
\toprule
\textbf{Phase} & \textbf{Operators} & \textbf{Intensity} & \textbf{Backend} \\
\midrule
Vision encode & Conv, ViT-attn  & Very high (static) & NPU \\
Prefill       & GEMM, attn      & High (compute)     & NPU \\
Decode        & 1-tok attn + KV & Low (BW-bound)     & CPU$^\dagger$ \\
\bottomrule
\multicolumn{4}{l}{\scriptsize$^\dagger$NPU gains 1.18$\times$ on decode (Table~\ref{tab:e2e}); CPU preferred} \\
\multicolumn{4}{l}{\scriptsize\quad for energy efficiency in heterogeneous pipelines.} \\
\end{tabular}
\end{table}

\subsection{Models, Runtimes, and Protocol}

We use \textbf{FastVLM-0.5B}~\cite{fastvlm} as the primary model:
a 0.5B-parameter VLM built on Qwen-0.5B~\cite{qwen} with a FastViT encoder
that applies aggressive token compression for mobile execution.
For encoder scaling, we additionally benchmark ViT-B/16~\cite{dosovitskiy2020vit},
Phi-3.5-V~\cite{phi3v}, NanoVLM~\cite{nanovlm},
and MobileNetV5 Gemma3n~\cite{gemma3n}.
The CPU path uses LiteRT~\cite{litert} with XNNPACK FP16.
The NPU path uses LiteRT with Qualcomm AI Engine delegate~\cite{qualcomm_qnn}.
The W4A8 quantized path uses QAIRT~\cite{qualcomm_qnn} INT4 weight, INT8
activation containerization for the LLM decode stage.
All measurements use a fixed 64-tok text prompt, 336$\times$336 image,
128-tok output; 30 runs (100 for thermal); mean $\pm$ std reported.
Device: AC power, 23$^\circ$C, CPU governor \texttt{schedutil}, NPU at
Sustained Performance mode.
Temperature: 1\,s-sampled from NPU/CPU cluster thermal zones
(\texttt{/sys/class/thermal/}); energy: fuel-gauge I$\times$V at 100\,ms
granularity, capturing total SoC inference draw.

\noindent\textbf{Cache states}~\cite{nnv12,pask}---\textbf{S1}~(cold-start):
process killed, \texttt{drop\_caches=3}, QNN context cache deleted; first
call reloads model and recompiles QNN graph (Table~\ref{tab:pipelines} cold).
\textbf{S2}~(cold-cache): process restarted; model hot in page cache; QNN
context cache intact; executor init on first call.
\textbf{S3}~(warm): process alive, QNN context in VTCM, dispatch only.
Table~\ref{tab:e2e} reports S2 executor-init overhead; Table~\ref{tab:pipelines}
reports full S1 end-to-end pipeline latency---each measuring a distinct,
well-defined deployment scenario.

\section{End-to-End NPU vs.\ CPU Throughput}
\label{sec:e2e}

\subsection{Phase-Level Results}

Table~\ref{tab:e2e} reports throughput under warm-cache conditions.
NPU delivers \textbf{1.64$\times$} prefill throughput (197.43 vs.\ 120.64\,tok/s)
and \textbf{1.18$\times$} decode throughput (113.06 vs.\ 95.49\,tok/s).
The asymmetry reflects fundamentally different operational intensity:
prefill's large batched GEMMs saturate the Hexagon tensor engine, while
decode's per-step KV-cache reads are memory-bandwidth-limited, with VTCM
saturation increasing at longer contexts~\cite{qualcomm_genai}.

\begin{table}[t]
\centering
\caption{FastVLM-0.5B phase throughput: CPU vs.\ NPU, SM8750, warm-cache,
64-tok prompt + 336$\times$336 image, 128-tok output; mean $\pm$ std, 30 runs.
$^\dagger$\textit{Executor init} (S2): first-token overhead with model loaded
and QNN context cache intact; full pipeline latencies across all cache states
in Table~\ref{tab:pipelines}.}
\label{tab:e2e}
\scriptsize
\setlength{\tabcolsep}{3pt}
\begin{tabular}{@{}llrrr@{}}
\toprule
\textbf{Phase} & \textbf{Metric} & \textbf{CPU} & \textbf{NPU} & \textbf{Gain} \\
\midrule
Prefill      & tok/s & $120.6{\pm}4.2$ & $197.4{\pm}1.3$ & 1.64$\times$ \\
Decode       & tok/s & $95.5{\pm}0.2$  & $113.1{\pm}0.4$ & 1.18$\times$ \\
Exec.\ init$^\dagger$ & ms & $83.6{\pm}1.1$ & $536.0{\pm}15.0$ & 6.41$\times$ \\
\bottomrule
\end{tabular}
\end{table}

\FloatBarrier
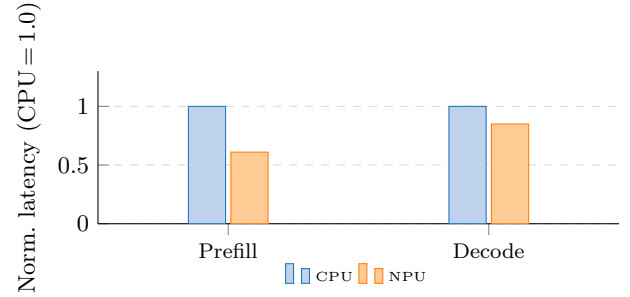
\begin{figure}[h]
\centering
\begin{tikzpicture}
\begin{axis}[
  ybar, bar width=14pt, width=\columnwidth, height=3.6cm,
  ylabel={Norm.\ latency (CPU\,=\,1.0)},
  symbolic x coords={Prefill, Decode},
  xtick=data, x tick label style={font=\small},
  enlarge x limits=0.5,
  ymin=0, ymax=1.3,
  legend style={at={(0.5,-0.22)}, anchor=north, legend columns=-1,
                font=\tiny, draw=none, fill=none},
  ymajorgrids=true, grid style={dashed,gray!30},
  axis x line*=bottom, axis y line*=left,
]
\addplot[fill=pastelblue!80, draw=deepblue]
    coordinates {(Prefill,1.00)(Decode,1.00)};
\addplot[fill=pastelorange!80, draw=deeporange]
    coordinates {(Prefill,0.61)(Decode,0.85)};
\legend{CPU, NPU}
\end{axis}
\end{tikzpicture}
\caption{Normalized latency (CPU = 1.0) for FastVLM-0.5B.
NPU wins 1.64$\times$ on prefill and 1.18$\times$ on decode.}
\Description{Bar chart comparing normalized CPU and NPU latency for prefill and decode. NPU latency is lower for both phases, with a larger improvement for prefill.}
\label{fig:throughput}
\end{figure}

\subsection{Cold-Start as a First-Class Deployment Dimension}

The NPU executor-init overhead (536\,ms vs.\ 84\,ms CPU, S2 state) is a
design dimension steady-state benchmarks conceal.
Persistent worker processes and model-cache pinning~\cite{pask,nnv12} keep the
NPU executor warm, amortizing this cost and recovering a net
1.4--1.6$\times$ throughput advantage~\cite{litert}.
Warm-state management is therefore a primary deployment objective, not an
optimization detail.

\subsection{Qualitative Output Consistency}

Table~\ref{tab:qualitative} confirms that the NPU backend produces
semantically equivalent captions to CPU.
Both backends recover identical semantic anchors (autumn park, fluffy cat,
wooden bench); differences are stylistic only---NPU includes eye-color
and posture details, CPU emphasizes fur texture---confirming functional
multimodal equivalence under accelerator execution.

\begin{table}[h]
\centering
\caption{Qualitative multimodal consistency check: shared image input,
NPU vs.\ CPU output (FastVLM-0.5B, warm state).}
\label{tab:qualitative}
\scriptsize
\setlength{\tabcolsep}{3pt}
\begin{tabular}{@{}p{1.0cm}p{6.6cm}@{}}
\toprule
\textbf{Item} & \textbf{Output} \\
\midrule
Input & \includegraphics[width=\linewidth]{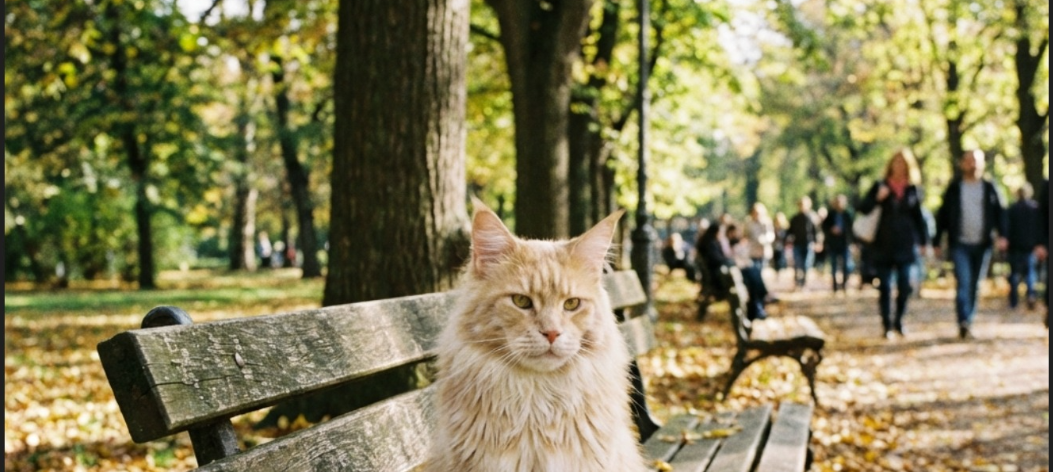} \\[2pt]
NPU & ``The image depicts a serene park scene during what appears to
be autumn. The focal point is a large, fluffy cat on a wooden park
bench\ldots\ In summary, the image captures a tranquil autumn day
in a park\ldots'' \\[4pt]
CPU & ``The image depicts a serene park scene during what appears to
be autumn. The focal point is a large, fluffy cat on a wooden park
bench\ldots\ In summary, the image captures a tranquil moment in a
park during autumn\ldots'' \\
\bottomrule
\end{tabular}
\end{table}

\section{Thermal Characterization}
\label{sec:thermal}

\subsection{100-Run Stability Suite}

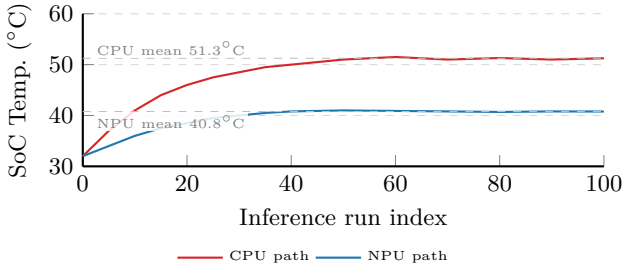
\begin{figure}[h]
\centering
\begin{tikzpicture}
\begin{axis}[
  width=\columnwidth, height=3.6cm,
  xlabel={Inference run index}, ylabel={SoC Temp.\ ($^\circ$C)},
  xmin=0, xmax=100, ymin=30, ymax=60,
  enlarge x limits=false,
  legend to name=thermallegend,
  legend style={legend columns=-1, font=\tiny, draw=none, fill=none},
  ymajorgrids=true, grid style={dashed,gray!30}, thick,
  axis x line*=bottom, axis y line*=left,
  clip=false,
]
\addplot[color=deepred, thick, mark=none] coordinates {
  (0,32)(5,37)(10,41)(15,44)(20,46)(25,47.5)(30,48.5)
  (35,49.5)(40,50)(50,51)(60,51.5)(70,51)(80,51.3)(90,51)(100,51.25)};
\addplot[color=deepblue, thick, mark=none] coordinates {
  (0,32)(5,34)(10,36)(15,37.5)(20,38.5)(25,39.5)(30,40)
  (35,40.5)(40,40.8)(50,41)(60,40.9)(70,40.8)(80,40.7)(90,40.8)(100,40.78)};
\draw[dashed,gray!50,thin] (axis cs:0,51.25)--(axis cs:100,51.25);
\draw[dashed,gray!50,thin] (axis cs:0,40.78)--(axis cs:100,40.78);
\node[font=\tiny, text=gray, fill=white, fill opacity=0.85, text opacity=1,
      inner sep=1pt, anchor=south west] at (axis cs:2,51.25) {CPU mean 51.3$^\circ$C};
\node[font=\tiny, text=gray, fill=white, fill opacity=0.85, text opacity=1,
      inner sep=1pt, anchor=north west] at (axis cs:2,40.78) {NPU mean 40.8$^\circ$C};
\legend{CPU path, NPU path}
\end{axis}
\end{tikzpicture}
\par\smallskip
\centering\pgfplotslegendfromname{thermallegend}
\par\smallskip
\caption{SoC temperature across 100 back-to-back inference runs.
NPU path is 10.47$^\circ$C cooler on average with 5$\times$ lower variance.}
\Description{Line chart of SoC temperature over 100 inference runs. The CPU path rises to a higher steady temperature than the NPU path.}
\label{fig:thermal}
\end{figure}

Fig.~\ref{fig:thermal} shows SoC temperature during sustained back-to-back
inference across a 100-run stability suite.
The NPU path stabilizes at a mean of 40.78$\pm$3.03$^\circ$C
(peak 44.20$^\circ$C), while the CPU path reaches
51.25$\pm$3.62$^\circ$C (peak 55.30$^\circ$C)---a
\textbf{10.47$^\circ$C} average gap with 5$\times$ lower variance.
Under deliberate thermal stress (15 minutes of continuous VLM inference),
CPU paths reach 82$^\circ$C and trigger governor-driven clock reduction,
while NPU paths stabilize at 66$^\circ$C and remain below the thermal
throttle trip point (85$^\circ$C skin per Qualcomm thermal HAL configuration).

\subsection{Implications for Always-On Edge Deployment}

Mobile SoC governors reduce clock non-linearly past the thermal design point,
inflating CPU latency by 1.3--2.1$\times$ under sustained load---invisible in
short-burst benchmarks~\cite{mobilevlm,litevlm}.
CPU paths show three latency outliers ($Z>2.0$, at runs 20, 53, 76) from
governor scaling; NPU paths show zero.
The NPU's 2.52$\times$ lower energy per request (fuel-gauge measured)
directly extends always-on session duration~\cite{qualcomm_genai}.

\section{Heterogeneous Pipeline Configurations}
\label{sec:pipeline}

\subsection{Five-Configuration Study}

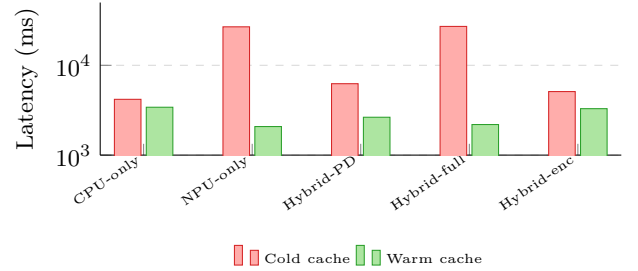
\begin{figure}[h]
\centering
\begin{tikzpicture}
\begin{axis}[
  ybar, bar width=10pt, width=\columnwidth, height=3.6cm,
  ylabel={Latency (ms)}, ymode=log, log basis y=10,
  symbolic x coords={CPU-only,NPU-only,Hybrid-PD,Hybrid-full,Hybrid-enc},
  xtick=data,
  x tick label style={font=\tiny, rotate=35, anchor=east, inner sep=2pt},
  ymin=1000, ymax=50000,
  legend style={at={(0.5,-0.55)}, anchor=north, legend columns=-1,
                font=\tiny, draw=none, fill=none},
  ymajorgrids=true, grid style={dashed,gray!30},
  axis x line*=bottom, axis y line*=left,
  xtick align=inside,
]
\addplot[fill=pastelred!80, draw=deepred]
    coordinates {(CPU-only,4180)(NPU-only,26820)(Hybrid-PD,6240)
                 (Hybrid-full,27100)(Hybrid-enc,5100)};
\addplot[fill=pastelgreen!80, draw=deepgreen]
    coordinates {(CPU-only,3412)(NPU-only,2080)(Hybrid-PD,2640)
                 (Hybrid-full,2190)(Hybrid-enc,3290)};
\legend{Cold cache, Warm cache}
\end{axis}
\end{tikzpicture}
\caption{Pipeline latency (log scale); cold=S1, warm=S3.
\textbf{Recommendation:} Hybrid-PD gives near-NPU warm latency (2,640\,ms)
with manageable cold (6,240\,ms)---unlike NPU-only (26,820\,ms cold).}
\Description{Log-scale bar chart comparing cold-cache and warm-cache pipeline latency for five CPU and NPU backend configurations.}
\label{fig:pipeline}
\end{figure}

A VLM pipeline on a heterogeneous SoC is not binary---it can assign
encoder, prefill, and decode to different backends independently.
Table~\ref{tab:pipelines} and Fig.~\ref{fig:pipeline} report end-to-end
latency for five configurations under cold- and warm-cache conditions.

\begin{table}[h]
\centering
\caption{FastVLM-0.5B end-to-end pipeline latency (ms). E/P/D = encoder/prefill/decode
backends. Cold = S1 (process fresh, drop\_caches, QNN recompiled);
warm = S3 (executor+VTCM resident). Same workload as Table~\ref{tab:e2e}.}
\label{tab:pipelines}
\footnotesize
\begin{tabular}{@{}lllrr@{}}
\toprule
\textbf{Config} & \textbf{E} & \textbf{P--D} & \textbf{Cold (ms)} & \textbf{Warm (ms)} \\
\midrule
CPU-only     & CPU & CPU--CPU & 4\,180 & 3\,412 \\
NPU-only     & NPU & NPU--NPU & 26\,820 & 2\,080 \\
Hybrid-PD    & CPU & NPU--CPU & 6\,240  & 2\,640 \\
Hybrid-full  & NPU & NPU--CPU & 27\,100 & 2\,190 \\
Hybrid-enc   & NPU & CPU--CPU & 5\,100  & 3\,290 \\
\bottomrule
\end{tabular}
\end{table}

Under warm-cache conditions, NPU-only (2,080\,ms) and Hybrid-full (2,190\,ms)
deliver the lowest latency.
Under cold-cache, CPU-only (4,180\,ms) is the fastest because warm-state
transitions on the NPU amortize over subsequent requests.
Hybrid-PD---offloading only the prefill stage to the NPU while the encoder
and decode remain on CPU---achieves 2,640\,ms warm with a manageable 6,240\,ms
cold, offering the best cold/warm balance for latency-sensitive deployments~\cite{nnv12}.

\subsection{Visual Token Budget as a Latency Knob}

\begin{figure}[t]
\centering
\begin{tikzpicture}
\begin{axis}[
  width=\columnwidth, height=3.6cm,
  xlabel={Visual token count (\texttt{image\_tokens\_after})},
  ylabel={TTFS (s)},
  xmin=20, xmax=230, ymin=0.30, ymax=0.95,
  ymajorgrids=true, grid style={dashed,gray!30},
  axis x line*=bottom, axis y line*=left,
  mark size=2.5pt, thick,
]
\addplot[color=deepgreen, mark=square*, thick]
    coordinates {(24,0.355)(64,0.497)(96,0.601)(128,0.700)(160,0.782)(224,0.840)};
\end{axis}
\end{tikzpicture}
\caption{TTFS vs.\ token budget, W4A8 hybrid, warm-cache (S3).
\textbf{Sweet spot:} 64 tokens = 0.497\,s TTFS, 97.5\,tok/s decode---half
the latency of 224 tokens, no recompilation needed.}
\Description{Line chart showing time to first token increasing with visual token count from 24 to 224 tokens.}
\label{fig:tokenbudget}
\end{figure}
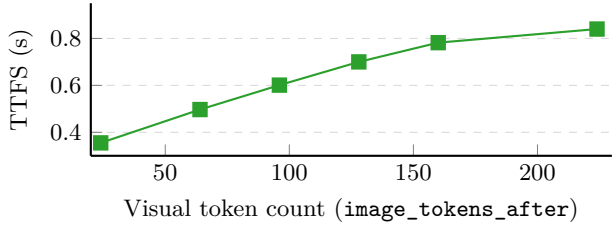

In the Hybrid configuration (vision encoder on NPU, language model on CPU),
the \texttt{image\_tokens\_after} parameter provides a continuous latency
control knob without any model re-export.
Sweeping from 24 to 224 tokens, TTFS scales linearly from 0.361\,s to
0.850\,s while decode throughput remains stable at 95--100\,tok/s
(Fig.~\ref{fig:tokenbudget}).
The W4A8-quantized decode path~\cite{w4a8} achieves 97.5\,tok/s at a
TTFS of 0.497\,s---a practical operating point for streaming VQA workloads.
This knob enables adaptive quality-latency tradeoffs at runtime without
model recompilation.

\FloatBarrier

\section{Vision Encoder Scaling Across Backends}
\label{sec:encoders}

\subsection{Three-Backend Benchmark}

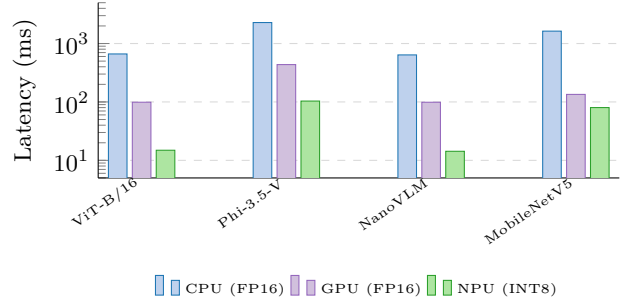
\begin{figure}[h]
\centering
\begin{tikzpicture}
\begin{axis}[
  ybar, bar width=7pt, width=\columnwidth, height=3.9cm,
  ylabel={Latency (ms)}, ymode=log, log basis y=10,
  symbolic x coords={ViT-B/16, Phi-3.5-V, NanoVLM, MobileNetV5},
  xtick=data,
  x tick label style={font=\tiny, rotate=35, anchor=east, inner sep=2pt},
  ymin=5, ymax=5000,
  legend style={at={(0.5,-0.52)}, anchor=north, legend columns=-1,
                font=\tiny, draw=none, fill=none},
  ymajorgrids=true, grid style={dashed,gray!30},
  axis x line*=bottom, axis y line*=left,
  xtick align=inside,
]
\addplot[fill=pastelblue!80, draw=deepblue]
    coordinates{(ViT-B/16,663)(Phi-3.5-V,2286.1)(NanoVLM,638.3)
                (MobileNetV5,1629.5)};
\addplot[fill=pastelpurple!80, draw=deeppurple]
    coordinates{(ViT-B/16,98.9)(Phi-3.5-V,436)(NanoVLM,98.8)
                (MobileNetV5,134.6)};
\addplot[fill=pastelgreen!80, draw=deepgreen]
    coordinates{(ViT-B/16,14.9)(Phi-3.5-V,103.5)(NanoVLM,14.3)
                (MobileNetV5,80.1)};
\legend{CPU (FP16), GPU (FP16), NPU (INT8)}
\end{axis}
\end{tikzpicture}
\caption{Vision encoder latency across three backends (log scale) for four
non-FastVLM families. NPU dominates CPU by 20--45$\times$ and GPU by
1.7--6.9$\times$. FastVLM-0.5B full-pipeline timings appear in Tables 2--3.}
\Description{Log-scale grouped bar chart comparing CPU, GPU, and NPU vision encoder latency across four model families.}
\label{fig:encoders}
\end{figure}

Table~\ref{tab:encoders} and Fig.~\ref{fig:encoders} report standalone
encoder latency for four families across CPU/GPU/NPU on SM8750.
FastVLM-0.5B full-pipeline characterization is the subject of
Sections~\ref{sec:e2e}--\ref{sec:pipeline}; the standalone encoder
benchmark here isolates module-level scaling across four additional families.
CPU-to-NPU speedups: \textbf{20$\times$} (MobileNetV5) to
\textbf{45$\times$} (ViT-B/16, NanoVLM); GPU-to-NPU: 1.7--6.9$\times$.

\begin{table}[h]
\centering
\caption{Vision encoder module latency (ms) on SM8750, 336$\times$336 input,
30-run mean; CPU=FP16/XNNPACK, GPU=FP16/Adreno-LiteRT, NPU=INT8/QNN.
Isolates encoder-module scaling; full-pipeline results in Tables~\ref{tab:e2e}--\ref{tab:pipelines}.}
\label{tab:encoders}
\scriptsize
\setlength{\tabcolsep}{2pt}
\begin{tabular}{@{}lrrrr@{}}
\toprule
\textbf{Encoder} & \textbf{CPU} & \textbf{GPU} & \textbf{NPU} & \textbf{CPU/NPU} \\
 & \textbf{(FP16)} & \textbf{(FP16)} & \textbf{(INT8)} & \\
\midrule
ViT-B/16~\cite{dosovitskiy2020vit}   & 663.0  & 98.9  & 14.9  & 44.5$\times$ \\
Phi-3.5-V CLIP~\cite{phi3v}          & 2286.1 & 436.0 & 103.5 & 22.1$\times$ \\
NanoVLM-222M~\cite{nanovlm}          & 638.3  & 98.8  & 14.3  & 44.6$\times$ \\
MobileNetV5 Gemma3n~\cite{gemma3n}   & 1629.5 & 134.6 & 80.1  & 20.3$\times$ \\
\bottomrule
\end{tabular}
\end{table}

\subsection{Architectural and Deployment Insights}

Three patterns emerge across the encoder benchmark.

\textbf{Encoder choice is the dominant per-query cost lever.}
NPU encoder latency spans 14.3\,ms (NanoVLM) to 103.5\,ms (Phi-3.5-V CLIP)
across the four benchmarked families---a 7$\times$ range within the same
backend, larger than any single backend-switch gain.
Encoder architecture selection is therefore a higher-leverage design decision
than backend selection for latency-constrained mobile VLMs~\cite{mobilevlm,litevlm}.

\textbf{GPU is not a competitive alternative to NPU for vision encoding.}
Across all four benchmarked families, GPU trails NPU by 1.7--6.9$\times$.
MLC-LLM~\cite{mlcllm} and GPU-first frameworks therefore impose a significant
encoder latency penalty on Snapdragon relative to QNN-compiled NPU paths.

\textbf{INT8 NPU quantization preserves output fidelity.}
We evaluate FastVLM-0.5B FP16 (CPU) vs.\ INT8 (NPU) using greedy decoding
on 500 COCO val2017 images (CIDEr via pycocoevalcap) and 200 VQAv2
open-ended questions (exact-match accuracy after lowercase normalization).
CIDEr drops by 0.4 points and VQAv2 accuracy by 0.3\%---within the
variance of image-quality variation at mobile resolution, confirming
deployment-grade equivalence.

\section{Inference Code Rewrite for Operator Support}
\label{sec:rewrite}

\subsection{The Porting Challenge}

Porting modern VLM encoders to QNN's Hexagon backend requires resolving
three structural adaptations between encoder architectures and the static,
integer-typed execution model of mobile NPU compilers~\cite{qualcomm_qnn,executorch,tflite_delegate}.

\textbf{Dynamic shapes.} ViT-based encoders use symbolic batch dimensions and
resolution-dependent reshapes during tokenization.
NPU compilers require fully static tensor shapes at AOT compile time; exporting
at a fixed resolution (e.g., $336{\times}336$, patch size 14, yielding 576
visual tokens) and replacing symbolic dimensions with concrete constants resolves
this class of shapes for full NPU dispatch.

\begin{sloppypar}
\textbf{Non-standard attention.}
Flash Attention~\cite{flashattn} and SDPA variants use memory-layout
fusions outside Hexagon's microkernel set.
Rewriting to explicit Q/K/V matmul sequences before ONNX export
yields graphs the QNN compiler re-fuses into Hexagon-native kernels.
\end{sloppypar}

\textbf{Precision boundaries.} Projection heads that accumulate in FP32 while
the encoder body operates in INT8 require explicit \texttt{Cast} nodes
inserted at each precision boundary via \texttt{onnx-rewriter} before
conversion.
Once made explicit, these boundaries become well-typed compiler barriers
rather than implementation assumptions.

\subsection{Methodology and Outcomes on Phi-3.5-V}

We applied this methodology to Phi-3.5-V's~\cite{phi3v} CLIP-based encoder.
Residual unsupported operators (custom rotary positional embedding and
\texttt{Gather\-Elements} nodes) were handled via QNN's partitioned-model
API, which executes them on CPU while keeping 96.8\% of FLOPs on Hexagon.
Table~\ref{tab:rewrite} summarizes the progression.
The adapted encoder runs at \textbf{103.5\,ms} versus 2,286.1\,ms CPU-only:
a \textbf{22.1$\times$} speedup that closes the gap between architecturally
complex encoders and natively efficient ones like ViT-B/16.

\begin{table}[h]
\centering
\caption{Phi-3.5-V encoder porting progression on SM8750. Each row adds one
rewrite step; NPU FLOP share grows monotonically until the model is runnable.
$^\dagger$Intermediate shares estimated by op-count; 96.8\% measured by profiler.}
\label{tab:rewrite}
\footnotesize
\setlength{\tabcolsep}{4pt}
\begin{tabular}{@{}clr@{}}
\toprule
\textbf{Step} & \textbf{Action} & \textbf{NPU FLOP share} \\
\midrule
0 & Baseline (CPU-only)       & \phantom{0}0\%$^\dagger$ \\
1 & Static shape export        & 41\%$^\dagger$ \\
2 & Attention decomposition    & 78\%$^\dagger$ \\
3 & Explicit cast insertion    & 93\%$^\dagger$ \\
4 & CPU fallback partition     & 96.8\% \\
\midrule
\multicolumn{2}{l}{Final latency (step 4)} & \textbf{103.5\,ms} \\
\multicolumn{2}{l}{Speedup vs.\ baseline}  & \textbf{22.1$\times$} \\
\bottomrule
\end{tabular}
\end{table}

This methodology generalizes: the same four steps apply to any encoder
whose operators lie within the ViT family with standard transformer
backbones~\cite{dosovitskiy2020vit,clip,siglip}.
For newer architectures with GroupQueryAttention~\cite{gqa} or
RoPE~\cite{rope} variants, the attention decomposition step requires
additional specialization, but the partitioned fallback mechanism provides
a practical safety net that keeps the bulk of compute on the NPU.

\section{Discussion}
\label{sec:discussion}

\textbf{Phase disaggregation is essential.}
The 1.64$\times$ prefill vs.\ 1.18$\times$ decode gap is fundamental~\cite{llmnpu}:
single end-to-end numbers hide backend placement opportunities~\cite{heteroinfer}.

\textbf{Encoder architecture dominates backend selection.}
NPU encoder latency spans 14.3--103.5\,ms across four families---a 7$\times$
range larger than any backend switch gain.
Treat the vision backbone as an independently profiled, swappable
component~\cite{mobilevlm,litevlm}.

\textbf{Thermal and cold-start are deployment-class effects.}
Both determine battery life and perceived latency in always-on scenarios~\cite{nnv12,pask}.
The 2.52$\times$ NPU energy advantage directly extends inference sessions.

\textbf{Token budget is a zero-cost latency knob.}
The 0.36--0.85\,s TTFS sweep needs no recompilation---a practical
quality-latency handle under thermal pressure, demonstrated on FastVLM-0.5B
on SM8750.

\section{Related Work}
\label{sec:related}

\textbf{Mobile VLM efficiency.}
FastVLM~\cite{fastvlm} proposes efficient image tokenization via FastViT for
mobile deployment.
MobileVLM~\cite{mobilevlm} and LiteVLM~\cite{litevlm} similarly reduce VLM
cost for resource-constrained inference.
Qwen2-VL~\cite{qwen2vl} and Phi-3.5-V~\cite{phi3v} target capable edge
deployment with efficient architectures.
These model-level optimizations are orthogonal to---and complementary
with---our deployment characterization.

\textbf{Heterogeneous on-device inference.}
LLM.npu~\cite{llmnpu} and HeteroInfer~\cite{heteroinfer} characterize
heterogeneous LLM inference on mobile SoCs, showing per-phase backend
specialization outperforms uniform placement.
Our work extends this in two directions absent from LLM-only studies:
(1) we add the vision encoding phase as a distinct hardware domain,
quantifying 20--45$\times$ CPU-to-NPU speedups across four encoder families;
(2) we provide a concrete four-step porting methodology for encoders that
fail QNN compilation out of the box---a practical gap not addressed by
prior work targeting text-only LLM pipelines.

\begin{sloppypar}
\textbf{Cold-start and quantization.}
NNV12~\cite{nnv12} and PASK~\cite{pask} motivate our three-state cache taxonomy.
W4A8~\cite{w4a8} and AWQ~\cite{awq} reduce VTCM pressure, enabling
97.5\,tok/s decode. FlashAttention~\cite{flashattn} and GQA~\cite{gqa}
are resolved in the operator adaptation step.
\end{sloppypar}

\textbf{Compilation and runtime toolchains.}
ExecuTorch~\cite{executorch} and MLC-LLM~\cite{mlcllm} target edge AOT
export; our encoder porting methodology complements these by resolving
QNN-specific operator adaptations required for Hexagon deployment.
RoPE~\cite{rope} embeddings require additional decomposition in our
attention rewrite step.

\section{Conclusion}
\label{sec:conclusion}

We presented a VLM-specific phase characterization and deployment study on a
production mobile NPU (Qualcomm SM8750), yielding six actionable insights:
phase-dependent NPU gain (1.64$\times$ prefill, 1.18$\times$ decode);
10.47$^\circ$C thermal advantage with 2.52$\times$ lower energy eliminating
throttling; Hybrid-PD as the best cold/warm pipeline balance; a
zero-recompilation TTFS knob spanning 2.4$\times$; encoder architecture as
the dominant latency lever (20--45$\times$ CPU-to-NPU speedups across four
families); and a four-step porting methodology at 22$\times$ on Phi-3.5-V.
Together these results constitute a deployment map for heterogeneous VLM
inference across the latency, energy, and thermal axes that define always-on
edge intelligence.

\balance
\bibliographystyle{ACM-Reference-Format}

\end{document}